\documentclass[12pt]{article}
\begin{document}
\begin{center}
{\Large Varying Light Velocity as a Solution to the Problems in Cosmology}
\vskip 0.3 true in
{\large J. W. Moffat}
\date{}
\vskip 0.3 true in
{\it Department of Physics, University of Toronto,
Toronto, Ontario M5S 1A7, Canada}

\end{center}

\begin{abstract}%
In earlier published work, it was proposed that light speed was larger in the early
universe by 30 orders of magnitude compared to its presently observed value. This
change in the speed of light is associated with a spontaneous breaking of local
Lorentz invariance in the early Universe, associated with a first order phase transition
at a critical time $t=t_c$. This solves the horizon problem, leads to a mechanism of
monopole suppression in cosmology and can resolve the flatness problem.
it also offers the potential of solving the cosmological constant
problem. After the critical time $t_c$, local Lorentz invariance
is restored and light travels at its presently measured speed. We
investigate the field equations in the spontaneously broken phase
and study further the flatness problem and the cosmological
constant problem. The entropy is shown to undergo a large
increase as the light velocity goes through a phase transition. A
scale invariant prediction for microwave background fluctuations
originating at the time of the phase transition is discussed.
\end{abstract}

\vskip 0.5 true in
UTPT-98. e-mail: moffat@medb.physics.utoronto.ca

\section{Introduction}

The idea that the velocity of light varies in the early Universe and several of its consequences
for cosmology was published some time ago\cite{Moffat1,Moffat2,Moffat3}. The idea
originated with the hypothesis that there is a phase of
spontaneously broken, local Lorentz invariance and diffeomorphism invariance due to a
non-vanishing vacuum expectation value (vev) of a field, $\phi$, shortly after the beginning
of the Universe\cite{Moffat3}. The local Lorentz and diffeomorphism symmetries of Einstein's
gravitational theory are spontaneously broken by the symmetry breaking pattern:
$SO(3,1)\rightarrow O(3)$ at a critical temperature
$T_c$, below which the symmetry is restored.  It was shown that this new scenario is
capable of solving the horizon problem, the excess relic particle and flatness problems
and leads to predictions for small scale inhomogeneities.

Recently, a series of papers, beginning with the paper by Albrecht and Magueijo[4-6],
has appeared in which the velocity of light (and possibly the gravitational constant
$G$ and other fundamental constants) was postulated to vary in the early universe.

In the following, we shall investigate further the physical
consequences of the spontaneous breaking of the symmetries of spacetime
in the early Universe, and the predictions for cosmology when the velocity of light
goes through a first order phase transition at a critical time $t\sim t_c$.

\section{Spontaneous Breaking of Spacetime Symmetries}

In the earlier work[1-3], we assumed that local Lorentz
vacuum symmetry is spontaneously broken by a Higgs mechanism.
We postulated the existence of four scalar fields, $\phi_a$, where $a$ labels the flat tangent
space coordinates, and assumed that the vev of the scalar fields, $<\phi_a>_0$, vanishes
for some temperature T less than a critical temperature $T_c$, when
the local Lorentz symmetry is restored. Above $T_c$ the non-zero
vev will break the symmetry of the gound state of the Universe from $SO(3,1)$
down to $O(3)$. The domain formed by the direction of the vev of the field $\phi$
will produce an arrow of time pointing in the direction of increasing entropy and the expansion
of the Universe.

Let us introduce the four real fields $\phi^a(x)$ (a=0,1,...3),
where $a,b,..$ denote flat tangent space coordinates. The
fields $\phi^a$ are scalars in coordinate space with the
coordinates $x^\mu$, and they are invariant
under Lorentz transformations
\begin{equation}
\phi^{\prime\,a}(x)=L^a_b(x)\phi^b(x).
\end{equation}
We can use a vierbein ${e^a}_\mu$ to convert $\phi^a$ into a 4-vector in
coordinate space: $\phi^{\mu}={e^\mu}_a\phi^a$.
The $e^a_\mu$ satisfy
\begin{equation}
e^a_\mu e^\mu_b=\delta^a_b,\quad e^\mu_a e^a_\nu=\delta^\mu_\nu,
\end{equation}
and they obey the Lorentz transformation rule
\begin{equation}
e^{'a}_\mu(x)=L^a_b(x)e^b_\mu(x).
\end{equation}
The covariant derivative operator acting on $\phi^a$ is defined by
\begin{equation}
D_{\mu}\phi^a=[\partial_{\mu}\delta^a_b+(\Omega_{\mu})^a_b]\phi^b,
\end{equation}
where $(\Omega_\mu)^a_b$ denotes the spin, gauge connection.

Consider the infinitesimal Lorentz transformation
\begin{equation}
L^a_b(x)=\delta^a_b+\omega^a_b(x)
\end{equation}
with
\begin{equation}
\omega_{ab}(x)=-\omega_{ba}(x).
\end{equation}
The matrix $D(L)$ in the transformation rule for a general field
$f_n(x)$ :
\begin{equation}
f_n(x)\rightarrow \sum_m[D(L)(x)]_{nm}f_m(x)
\end{equation}
takes the form
\begin{equation}
D[1+\omega(x)]=1+\frac{1}{2}\omega^{ab}(x)\sigma_{ab},
\end{equation}
where the $\sigma_{ab}$ are the six generators of the Lorentz
group which satisfy $\sigma_{ab}=-\sigma_{ba}$ and the
commutation rules
\begin{equation}
[\sigma_{ab},\sigma_{cd}]=\eta_{cb}\sigma_{ad}-\eta_{ca}\sigma_{bd}+\eta_{db}\sigma_{ca}
-\eta_{da}\sigma_{cb}.
\end{equation}
The set of scalar fields $\phi$ transforms as
\begin{equation}
\phi'(x)=\phi(x)+\omega^{ab}(x)\sigma_{ab}\phi(x).
\end{equation}
The gauge spin connection which satisfies the transformation
law\begin{equation}
(\Omega_\sigma)^a_b\rightarrow
[L\Omega_\sigma L^{-1}-(\partial_\sigma L)L^{-1}]^a_b,
\end{equation}
is given by
\begin{equation}
\Omega_\mu=\frac{1}{2}\sigma^{ab}e^\nu_ae_{b\nu;\mu},
\end{equation}
where $;$ denotes covariant differentiation with respect to the
Christoffel symbol $\Gamma^\lambda_{\mu\nu}$:
\begin{equation}
\Gamma^\lambda_{\mu\nu}=g^{\lambda\rho}\eta_{ab}(D_\mu
e^a_\nu)e^b_\rho.
\end{equation}

A Higgs sector is included in the Lagrangian density such
that the gravitational vacuum symmetry, which is set equal to the Lagrangian
symmetry at low temperatures, breaks to a smaller symmetry at high
temperature. The pattern of vacuum phase transition that emerges contains
a symmetry anti-restoration. This vacuum symmetry breaking leads
to the interesting possibility that exact zero temperature conservation laws
e.g. electric charge and baryon number are broken in the early
Universe. It was shown that the spontaneous breaking of the
Lorentz symmetry of the vacuum leads to a violation of the exact zero
temperature conservation of energy in the early Universe, which can explain
the origin of matter in the big bang.

Let us consider the Lorentz invariant potential:
\begin{equation}
V(\phi)=-{1\over 2}\mu^2\sum_{a=0}^3\phi_a\phi^a+\lambda\sum_{a=0}^3
(\phi_a\phi^a)^2,
\end{equation}
where we choose $\phi^a$ to be a timelike Lorentz vector,
$\phi_a\phi^a > 0$, and $\lambda > 0$, so that the potential is
bounded from below. If $V$ has a minimum at $\phi_a=v_a$, then
the spontaneously broken solution is given by
$v_a^2=\mu^2/4\lambda$ and an expansion of $V$ around the minimum
yields the mass matrix: \begin{equation} \label{matrix}
(\mu^2)_{ab}={1\over 2}\biggl({\partial^2 V\over \partial \phi_a \partial
\phi_b}\biggr)_{\phi_a=v_a}.
\end{equation}
We can choose $\phi_a$ to be of the form
\begin{equation}
\phi_a=\left(\matrix{0\cr 0\cr 0\cr v\cr}\right)
=\delta_{a0}(\mu^2/4\lambda)^{1/2}.
\end{equation}

All the other solutions of $\phi_a$ are related to this one by a Lorentz
transformation. Then, the homogeneous Lorentz group $SO(3,1)$ is broken
down to the spatial rotation group $O(3)$. The three rotation generators $J_i
(i=1,2,3)$ leave the vacuum invariant
\begin{equation}
J_iv_i=0,
\end{equation}
while the three Lorentz-boost generators $K_i$ break the vacuum symmetry
\begin{equation}
K_iv_i\not= 0.
\end{equation}

The mass matrix $(\mu^2)_{ab}$ can be calculated from (\ref{matrix}):
\begin{equation}
(\mu^2)_{ab}=(-{1\over 2}\mu^2+2\lambda v^2)\delta_{ab}+4\lambda v_av_b
=\mu^2\delta_{a0}\delta_{b0},
\end{equation}
where $v$ denotes the magnitude of $v_a$. There are three zero-mass Goldstone
bosons, the same as the number of massive vector bosons, $V^i_\mu=(\Omega_\mu)_{0i}
=-(\Omega_\mu)_{i0}$, and there are three
massless vector bosons, $U_\mu^n=(\Omega_\mu)_{mn}=-(\Omega_\mu)_{nm}$,
corresponding to the unbroken $O(3)$ symmetry. In
addition to these particles, one massive physical boson particle
$h$ remains, after the spontaneous breaking of the vacuum.

A phase transition is assumed to occur at the critical temperature $T_c$,
when $v_a\not= 0$ and the Lorentz symmetry is broken and the three gauge
fields $(\Omega_{\mu})_{i0}$ become massive vector bosons. Below
$T_c$ the Lorentz symmetry is restored, and we
regain the usual classical gravitational field with massless gauge fields
$\Omega_{\mu}$. The symmetry breaking will extend to the
singularity or the possible singularity-free initial state of the big bang,
and since quantum effects associated with gravity do not become important
before $T\sim 10^{19}$ GeV, we expect that $T_c\leq 10^{19}$ GeV.

The total action for the theory is
\begin{equation}
S=S_G+S_M+S_{\phi},
\end{equation}
where the action for Einstein gravity is
\begin{equation}
S_G=-{c^4\over 16\pi G}\int d^4x e(R+2\Lambda).
\end{equation}
$e\equiv \sqrt{-g}=\hbox{det}(e^a_{\mu}e_{a\nu})^{1/2}$, $\Lambda$
is the cosmological constant and $S_M$ is the matter action for gravity. Moreover,
\begin{equation}
S_{\phi}=\int d^4x\sqrt{-g}[{1\over 2}D_{\mu}\phi_a
D^{\mu}\phi^a-V(\phi)].
\end{equation}
By choosing $\phi^a$ to be a Lorentz timelike vector, we ensure
that the kinetic energy term $D_\mu\phi_aD^\mu\phi^a > 0$
for all events in the past and future light cones of the flat
tangent space. Since the kinetic energy term is positive definite
within the light cone and $\phi_a\phi^a > 0$, we guarantee
that local Lorentz invariance is broken within the past and
future light cones where spacetime events are causally connected.

Let us consider small oscillations about the true minimum and define a shifted field
\begin{equation}
\phi^{\prime}_a=\phi_a-v_a.
\end{equation}
We perform a Lorentz transformation on $\phi^a$, so that we obtain
\begin{equation}
\label{specialframe}
\phi^0=h,\quad \phi^1=\phi^2=\phi^3=0.
\end{equation}
In this special coordinate frame, the remaining component $h$ is the physical Higgs
particle that survives after the three Goldstone modes have been removed. This
corresponds to choosing the ``unitary gauge'' in the standard electroweak
model.

The total action for the theory in the broken symmetry phase, $T>T_c$, is
\begin{equation}
S=S_G+S_M+S_h+S_V.
\end{equation}
In our specially chosen coordinate frame in which (\ref{specialframe}) holds, we have
\begin{equation}
S_h=\int d^4x\sqrt{-g}[{1\over 2}\partial_{\mu}h
\partial^{\mu}h-V(h)],
\end{equation}
where
\begin{equation}
V(h)=4\lambda v^2h^2+4\lambda vh^3+\lambda h^4-{1\over 2}V_\mu^2h^2-vV_\mu^2h,
\end{equation}
and we have for convenience suppressed the index $i$ on $V_\mu^i$. Moreover,
\begin{equation}
S_V={1\over 2}m^2\int d^4x\sqrt{-g}g^{\mu\nu}V_{\mu}V_{\nu},
\end{equation}
where the mass $m\propto <h>_0$.

The field equations are of the form:
\begin{equation}
\label{Einstein}
G^{\mu\nu}\equiv R^{\mu\nu}-{1\over 2}g^{\mu\nu}R={8\pi G\over c^4}
(T^{\mu\nu}+K^{\mu\nu}+H^{\mu\nu})+\Lambda g^{\mu\nu},
\end{equation}
where $K^{\mu\nu}$ is given by
\begin{equation}
K^{\mu\nu}=m^2(V^{\mu}V^{\nu}-{1\over 2}g^{\mu\nu}V^{\beta}V_{\beta}).
\end{equation}
Moreover, the $h$ field energy-momentum tensor is of the usual form:
\begin{equation}
H^{\mu\nu}=\partial^{\mu}h \partial^{\nu}h - {\cal L}_h g^{\mu\nu}.
\end{equation}

Since $G^{\mu\nu}$ satisfies the Bianchi identities:
\begin{equation}
{G^{\mu\nu}}_{;\nu}=0,
\end{equation}
we find from (\ref{Einstein}) that
\begin{equation}
{T^{\mu\nu}}_{;\nu}=-(K^{\mu\nu}+H^{\mu\nu})_{;\nu},
\end{equation}
where we have used ${g^{\mu\nu}}_{;\nu}=0$.
In the unbroken phase of spacetime, we regain the standard energy-momentum
conservation laws ($K^{\mu\nu}=0,\, {H^{\mu\nu}}_{;\nu}=0$):
\begin{equation}
{T^{\mu\nu}}_{;\nu}=0,
\end{equation}
and the spin connection corresponds to a massless graviton gauge field.
 
\section {Field Equations in the Broken Symmetry Phase}

The spacetime manifold in the broken phase has the symmetry
$R\times O(3)$. The three-dimensional space with $O(3)$ symmetry
is assumed to be homogeneous and isotropic and yields the usual
maximally symmetric three-dimensional space:
\begin{equation}
d\sigma^2=R^2(t)\biggl[{dr^2\over 1-kr^2}+r^2(d\theta^2
+\hbox{sin}^2\theta d\phi^2)\biggr],
\end{equation}
where $t$ is the external
time variable. This is the Robertson-Walker theorem for our ordered phase
of the vacuum and it has the correct subspace structure for the FRW
Universe with the metric:
\begin{equation}
\label{metric}
ds^2=dt^2c^2-R^2(t)\biggl[{dr^2\over 1-kr^2}+r^2(d\theta^2
+\hbox{sin}^2\theta d\phi^2)\biggr].
\end{equation}

In the broken symmetry phase, the ``time" $t$ is the {\it absolute physical
time} measured by standard clocks. In contrast to GR, while $<\phi>_0$ is
non-zero, we no longer have re-parameterization invariance and time is no
longer an arbitrary label.

Let us consider Einstein's field equations in the broken symmetry phase $T > T_c$. We have
$g_{00}(t)=c_0^2$ and $g_{ik}(t)=-R^2(t)\gamma_{ik}$ and the energy-momentum tensor for a
perfect fluid takes the form:
\begin{equation}
\label{energy}
T^{\mu\nu}=\biggl(\rho+\frac{p}{c_0^2}\biggr)u^\mu u^\nu-pg^{\mu\nu},
\end{equation}
where $c_0$ labels the velocity of light in this epoch, $u^\mu=dx^\mu/ds$, $u^\mu
u_\mu=c_0^2$, $\rho$ is the
density of matter and radiation and $p$ is the pressure. In our
homogeneous space the spatial part of the massive vector
field $V^i=0$. Let us use the notation: $V^0(t)=\chi(t)$. We
obtain the field equations in the broken symmetry phase
\begin{equation}
\label{Friedmann}
\frac{\dot R^2}{c_0^2R^2}+\frac{k}{R^2}=\frac{8\pi G}{3c_0^2}\rho
+\frac{8\pi G}{c_0^4}\biggl[\frac{1}{2}m^2\chi^2+\frac{1}{2}(\dot h)^2
+V(h)\biggr]+\frac{\Lambda}{3},
\end{equation}
\begin{equation}
2\biggl(\frac{\ddot R}{c_0^2R}\biggr)+\frac{\dot R^2}{c_0^2R^2}+\frac{k}{R^2}=
-\frac{8\pi G}{c_0^4}\biggl[p+\frac{1}{2}m^2\chi^2+\frac{1}{2}(\dot
h)^2-V(h)\biggr]+\Lambda,
\end{equation}
\begin{equation}
\frac{1}{R^3}\frac{\partial}{\partial t}
\biggl[R^3c_0^2\biggl(\rho+\frac{p}{c_0^2}\biggr)\biggr]-{\dot p}=-W,
\end{equation}
where $\dot R=dR/dt$ and $W$ is given by
\begin{equation}
W=m^2\chi\biggl[\dot\chi+3\biggl(\frac{\dot
R}{R}\biggr)\chi\biggr]+(\dot h)^2 \biggl[1+3\biggl(\frac{\dot
R}{R}\biggr)\bigg]+V^{\prime}(h){\dot h}.
\end{equation}

\section{Superluminary Universe}

The horizon scale is determined by
\begin{equation}
\label{scale}
d_H(t)=c_0R(t)\int_0^t{dt^{\prime}\over R(t^{\prime})}.
\end{equation}
For $t > t_c$, this will have the usual value: $d_H(t) = 2ct$,
since $R(t)\propto t^{1/2}$ for a radiation dominated Universe.
Let us assume that for $t \leq t_c$, the speed of light is very
large. During a first order phase transition, the velocity of light is
assumed to undergo a discontinuous change from the value:
\begin{equation}
\label{cconstant}
c_0\sim ac
\end{equation}
for $t\leq t_c$ to $c_0=c$ (c is the present value of the velocity of
light and $a$ is a constant) for $t > t_c$. Then, we get for $t\leq t_c$:
\begin{equation}
d_H(t)\approx c_0g(t),
\end{equation}
where $g(t)$ is the dynamical time dependence arising from $R(t)$ in
(\ref{scale}). Thus, for a fraction of time near the beginning of the Universe, and for
$a\rightarrow\infty$, all points of the expanding space will have been in communication with
one another solving the horizon problem.

Suppose the field $\phi$ that breaks the spacetime symmetries is characterized
by a correlation length $\xi$. Then, the monopole density is approximately
given by
\begin{equation}
n_M\approx \xi^{-3}.
\end{equation}
In the superluminary model, the bound on the length $\xi$ is given by
\begin{equation}
\xi < d_H(t) \approx c_0g(t),
\end{equation}
so that the bound on the number density of monopoles is exponentially
weakened. This solves the relic particle (monopole) problem.

The present observational data restrict
$\Omega_0=\rho_{\rm crit}/\rho_0$ to lie in the interval [0.01, few], which implies that
$R_{\rm curv}\sim c/H_0$ and $\rho_0\sim \rho_{\hbox{crit}}$. From Eq.(\ref{Friedmann})
in the broken phase, we can derive the expression
\begin{equation}
\Omega(t)=1/[1-x(t)],
\end{equation}
where, in the radiation dominated superluminary era,
\begin{equation}
x(t)=\frac{c_0^2k}{R^2H^2}\sim \frac{c_0^2k/R^2}{8\pi G\rho_r/3},
\end{equation}
where $\rho_r$ denotes the radiation density. Moreover we have
\begin{equation}
\Omega=\Omega_{\rho_r}+\Omega_F+\Omega_{\Lambda},
\end{equation}
in which
\begin{equation}
\Omega_{\rho_r}=\frac{8\pi G\rho_r}{3H^2},\quad \Omega_F
=\biggl(\frac{8\pi G}{c_0^2H^2}\biggr)\biggl[\frac{1}{2}m^2\chi^2+\frac{1}{2}(\dot h)^2
+V(h)\biggr],
\end{equation}
\begin{equation}
\Omega_{\Lambda}=\frac{c_0^2\Lambda}{3H^2}.
\end{equation}

We have in the radiation dominated era, $\rho_r=\rho_{0r}\biggl(R_0/R\biggr)^4$, so that
\begin{equation}
x\sim \frac{c_0^2kR^2}{R^*},
\end{equation}
where $R^*=8\pi G\rho_{0r}R_0^4/3$.
This yields
\begin{equation}
\vert\Omega(10^{-43}{\rm sec})-1\vert\sim O(a^210^{-60}).
\end{equation}
Thus, in the short time that the Universe is superluminary with
$a\sim 5\times 10^{29}$, we get
\begin{equation}
\vert\Omega(10^{-43}{\rm sec})-1\vert\sim O(1),
\end{equation}
which implies much less fine tuning than the standard FRW model.

We observe that in contrast to
the inflationary model (which predicts that $\Omega=1$)\cite{Guth,Turner}, the prediction for
the value of $\Omega$ in the superluminary model depends on the detailed dynamics
of the theory\cite{Moffat1,Albrecht,Barrow1,Barrow2}. Indeed, if we were to assume the
equation of state: $\rho=$ const., that $\chi$ and $V$ are uniform in the broken phase and
$k=0$, then $R(t)$ has the inflationary solution:
\begin{equation}
R(t)\propto \hbox{exp}\biggl[tc_0\biggl(\Lambda/3\biggr)^{1/2}\biggr].
\end{equation}
Thus we would regain the standard inflationary prediction
$\Omega=1$. Clearly, the superluminary model {\it does not lead automatically to the generic
prediction} $\Omega=1$. The possibility of obtaining an open universe version of the
inflationary scenario has been the subject of much controversy recently
\cite{Hawking,Linde,Vilenkin}. The fact that obtaining an open Universe in the superluminary
model is not a problem is a positive feature in favour of the model.

Let us now consider the cosmological constant problem. We ignore
the effects of the $h$ and $V$ fields, since they will not play
an important role in the present discussion. Then, using the
equation of state: $p=\frac{1}{3}\rho$, we can derive at some
instant of time ${\bar t}$: \begin{equation}
H^2({\bar t})(q({\bar t})-1)
=\frac{c_0^2k}{R^2({\bar t})}-\frac{2\Lambda c_0^2}{3},
\end{equation}
where $q$ is the deceleration parameter
\begin{equation}
q=-{\ddot R}R/{\dot R}^2.
\end{equation}
We also have
\begin{equation}
\frac{c_0^2k}{R^2({\bar t})H^2({\bar t})}
=\Omega({\bar t})-1+\frac{c_0^2\Lambda}{3H^2({\bar t})}.
\end{equation}
It follows that
\begin{equation}
\vert q({\bar t})-\Omega({\bar t})\vert=\vert
\frac{c_0^2\Lambda}{3H^2({\bar t})}\vert.
\end{equation}

Assuming that the radiation dominant solution of the field equations holds near the phase
transition, then  $\Omega({\bar t})\sim 0.1-1$, $q({\bar t})\sim
1$, $H({\bar t})\sim 1/2{\bar t}$ and
\begin{equation}
R({\bar t})=\biggl(\frac{32\pi
G\rho_{0r}}{3}\biggr)^{1/4}R_0{\bar t}^{1/2}.
\end{equation}
We obtain
\begin{equation}
\label{Lambdaeq}
\vert\Lambda \vert\approx \frac{1}{c_0^2{\bar t}^2}.
\end{equation}
For a rapid phase transition in the velocity of light
and for $c_0=ac$ and ${\rm log}_{10}a\geq 60$, we have for
${\bar t}\sim 10^{-43}\,{\rm sec}$:
\begin{equation}
\label{bound}
\vert\Lambda\vert < 10^{-54}\,{\rm cm}^{-2}.
\end{equation}
From the critical density $\rho_{\rm crit}\sim\rho_0$ this bound corresponds to
$\Lambda/8\pi G\leq 8\times 10^{-47}\,h^2\, {\rm GeV}^4$, where $0.4\leq h\leq 1$. This
would solve the cosmological constant problem\cite{Albrecht,Barrow1,Moffat2}. The
observational bound in (\ref{bound}) is obtained by using $H_0\sim 100\,{\rm km}
\,{\rm s}^{-1}\,{\rm Mpc}^{-1}$.

However, there is a serious conflict between the value ${\rm log}_{10}a\leq 30$ required
to solve the flatness problem and the value ${\rm log}_{10}a > 60$ required to solve the
cosmological constant problem. Indeed, the latter value is far too large to accomodate a
reasonable evolution of the FRW Universe just after the occurrence of the phase transition
in the velocity of light. However, if we assume that another
phase transition in the velocity of light occurs, before the one
that solves the flatness problem, with ${\rm log}_{10}\sim 60$,
then this could solve the cosmological constant problem and be
followed by a phase transition with a lowering of the speed of
light to a value with ${\rm log}_{10}\sim 30$, which could solve
the flatness problem and allow the universe to expand to its
present day value.

We have assumed the radiation dominant solution for $R(t)$ in this
derivation. Perhaps another dynamical solution of the field equations would accomodate a
solution to the horizon, flatness and cosmological constant problems in the presence
of a phase transition in the velocity of light. Barrow\cite{Barrow1}
has considered the time dependent solutions of Brans-Jordan-Dicke theories with a field
$\psi=c^4$, but such theories are severely restricted dynamically and can easily lead to
consistency problems.  In any eventuality, the horizon, flatness and monopole problems can be
resolved by the superluminary model. The problem with the
cosmological constant is not resolved in inflationary models, and
indeed is exacerbated by the enormous vacuum energy density
required to drive the initial inflation. The potential for the
superluminary model to solve the cosmological constant problem
could provide the model with another significant advantage.

One important aspect of the superluminary model is that it is
not sensitive to the choice of an equation of state. No exotic
forms of matter withe negative pressure are required to resolve
cosmological problems. This is an advantage over the standard
inflationary scenarios which require vacuum energy with $p=-\rho$
and unusual forms of potentials for the inflaton field that often
require fine-tuning to implement the inflationary period.

\section{Black-Body Radiation and Entropy in the Superluminary Phase}

In the Lorentz symmetry broken phase of the Universe, the total blackbody radiation energy at
temperature $T$ is
given by
\begin{equation}
u=\biggl(\frac{8\pi
h\nu^3}{c_0^3}\biggr)\int_0^{\infty}[\exp\biggl(\frac{h\nu}{kT}\biggr)-1]^{-1}d\nu
=\frac{8\pi^5(kT)^4}{15h^3c_0^3}.
\end{equation}
The number density of photons is
\begin{equation}
n_{\gamma}= \frac{60.42198(kT)^3}{(hc_0)^3}.
\end{equation}
The energy densities for photons and neutrinos are
\begin{equation}
\epsilon_\gamma=\sigma_B T^4,\quad \epsilon_\nu=\frac{7}{16}\sigma_B T^4,
\end{equation}
where $\sigma_B$ is the Stefan-Boltzmann constant
\begin{equation}
\sigma_B=\frac{\pi^2k_B^4}{60\hbar^3c_0^3}
\end{equation}
and $k_B$ is Boltzmann's constant.  For ${\rm log}_{10}a \leq 30$ Stefan-Boltzmann's
constant is $\sigma_B\sim 1.7\times 10^{-84}\,{\rm erg}\,{\rm cm}^{-3}\, K^{-4}$ in the
superluminary phase.

The Universe has negligible thermal and neutrino energy during the short period of the
spontaneously broken symmetry phase. After the superluminary phase ends when
$c_0=c$, both $\epsilon_\gamma$ and $\epsilon_\nu$ regain their standard values in an FRW
Universe. In contrast to the inflationary model, there is no problem with a reheating epoch
necessary in the inflationary scenario to replenish the matter and radiation in an `empty'
de-Sitter Universe.  The velocity of light phase transition in the spontaneous symmetry
breaking process automatically takes care of the creation of matter when
$c_0=c$\cite{Mak1,Mak2}.

The entropy of thermal photons is
\begin{equation}
S(T)=\frac{4\pi^2V}{45}\biggl(\frac{T}{\hbar c_0}\biggr)^3.
\end{equation}
Thus, at the phase transition at $t\sim t_c$, the entropy increases enormously in the direction of
the expanding universe when $c_0=c$. The arrow of time connected with the increase of the
entropy is determined by the domain arrow produced by
the non-zero $<\phi>_0$ in the spontaneously broken phase [1-3].

We see that the large value of $c$ in the early Universe, in the broken symmetry phase, changes
radically the thermal physics at the beginning of the universe. This is in accord with the known
observation that the entropy of a system increases rapidly as the system undergoes a first order
or second order phase transition from an ordered to a more disordered state.

\section{Quantum Fluctuations and Density Perturbations}

Let us consider the possibility in our model of generating the seed
perturbations that can grow to form the large-scale structures.
During the superluminary phase for $t < t_c$, the fluctuation wavelengths grow as
$\lambda\propto R(t)$. However, the horizon grows rapidly, $d_H\approx c_0g(t)$,
where $c_0$ is given by (\ref{cconstant}), and it will become equal to the physical
wavelength at some time $t=t_{\hbox{exit}}$, after which it becomes larger than
$\lambda(k)$ for a mode labeled by a wave vector ${\vec k}$.

After the symmetry is restored at $t > t_c$, the proper length $R(t)$ grows as $t^{1/2}$,
whereas the horizon will increase as $cH(t)^{-1}\sim ct$. Therefore, the
wavelength will be completely within the Hubble radius for an interval of time
$\Delta t$.  Thus, in the superluminary model the fluctuations are
in microcausal connection very early in the Universe ($t\sim 10^{-35}$ s)
and have time to grow into physical modes sufficiently large to form Galaxy structures.
These fluctuations will have a gaussian form, provided any self-couplings of the
matter fields are small.

The fluctuations associated with the Higgs field, $h$, could be a candidate
for seed perturbations. The $h$ field satisfies
\begin{equation}
\ddot h+3H\dot h+V^{\prime}(h)=0,
\end{equation}
where $V'(h)=dV(h)/dh$.

When the velocity of light undergoes a discontinuous
change, during a first order phase transition at $t\sim t_c$, to the value $c_0$ given by
(\ref{cconstant}) with $a\geq 10^{30}$,
then the horizon, $d_H(t)$, determined by (\ref{scale}) will also
have a discontinuity in its first derivative with respect to $t$,
and $d_H(t)$ for $t < t_c$ can be matched to $d_H(t)$ for $t \geq
t_c$ in such a way that $\lambda$ crosses $d_H(t)$ twice. The
fluctuations are ``frozen in'' and leave an imprint on the metric
tensor.

Fluctuations in $h$ give rise to perturbations in the density
\begin{equation}
\delta \rho_h=\delta h\biggl({\partial V\over \partial h}\biggr).
\end{equation}
At horizon crossings, $\lambda_{\hbox{phys}}\sim cH^{-1}$, the gauge invariant
quantity $\zeta$ takes the simple form
$\zeta=\delta\rho/(\rho+p/c^2)\,$\cite{Bardeen}. In the radiation
dominated era and in the matter dominated era, $\zeta$ at horizon
crossing is, up to a factor of order unity, equal to $\delta
\rho/\rho$. Equating the values of $\zeta$ at the two horizon
crossings, we find
\begin{equation}
\label{delta}
\biggl({\delta \rho\over \rho}\biggr)_{\hbox{Hor}}
\sim {\delta h V^{\prime}\over \dot h^2}\sim
{HV^{\prime}\over 2\pi \dot h^2},
\end{equation}
where we have used the fact that $\delta h\sim H/2\pi$.
We must now model $V^{\prime}$ and $\dot h$ at the phase transition, in order
to estimate the density fluctuation, $\delta \rho/\rho$. Clearly, $H$ is
rapidly varying at the phase transition. We have
\begin{equation}
\label{hdot}
\dot h\propto {H\over 2\pi \delta t}.
\end{equation}
A natural time scale for the duration of the phase transition is given by
\begin{equation}
\label{deltat}
\delta t\propto\biggl({H\over 2\pi h^3}\biggr)^{1/2}.
\end{equation}
Thus, if we choose $1/H\sim 10^{-34}$ s and $h\sim M_P\propto 10^{43} s^{-1}$,
then the duration of the phase transition is $\delta t\sim 10^{-48}$ s.

By assuming that $V(h)$ is dominated by $V(h)
\sim {\lambda\over 4} h^4$, we obtain from Eqs. (\ref{delta}),
(\ref{hdot}) and (\ref{deltat}) the scale invariant prediction
for the amplitude
\begin{equation}
\biggl({\delta\rho\over \rho}\biggr)_{\hbox{Hor}}\sim \lambda.
\end{equation}
We can fit the data measured by
the Cosmic Background Explorer (COBE), which is consistent with a gaussian,
scale invariant spectrum\cite{Harrison,Smoot}, by choosing the coupling constant
$\lambda\sim 10^{-5}$, and using
$\Delta T/T\sim {1\over 3}\delta\rho/\rho$. The measurements are quoted in terms
of a spectral index $n$, with $n=1.1\pm 0.5$. These measurements are
also consistent with the predictions of inflationary models, and with other
mechanisms of inhomogeneity generation, such as cosmic strings.

\section{Conclusions}

The results obtained above suggest that the superluminary model could be an attractive
alternative to inflation as a solution to the initial value problem in cosmology. Moreover, our
picture of the period immediately following the big bang is radically altered from the standard
big bang model. The violation of the conservation of energy in the spontaneously broken
symmetry phase can provide an explanation for the creation of matter in the beginning of the
Universe, an explanation which is not available in the standard FRW model or the inflationary
scenario. The model also provides a possible solution the cosmological costant problem.

The quantum fluctuations of the Higgs field, near the phase transition in the velocity of light,
produce microwave background density fluctuations, which are frozen in at the horizon. The
spectrum of the fluctuations is gaussian and scale invariant with a scalar field coupling constant,
$\lambda \sim 10^{-5}$, which is more reasonable in size than the standard value,
$\lambda\sim 10^{-14}$, predicted by generic inflationary models\cite{Guth,Turner}. Further
work is necessary to investigate in more detail the predictions of the density fluctuation
spectrum in the superluminary model.

The superluminary model of the early Universe[1-3] was introduced to provide an interesting
alternative to the inflationary model, at a time when the latter model was enjoying a popular
revival. It is remarkable that that there has been no serious alternative model considered
besides the superluminary model (or varying light speed model, as it is called by Albrecht and
Maguiejo\cite{Albrecht}). With the advent of a new generation of accurate satellite
measurements of the microwave background, it may be possible to distinguish the specific
predictions of the two models and the predictions of other alternative models that may be
forthcoming in the future.

An important feature of the superluminary model is the unavoidable breaking of
local Lorentz invariance in the early Universe. In view of
the significant changes that will occur in the fundamental physics, it is
necessary to have a well defined model for this symmetry breaking. Such a
model is provided by the spontaneous symmetry breaking of the Lagrangian in
the scenario presented in earlier publications[1-3] and in the present
exposition. The ``hidden" symmetry of the gravitational vacuum has the advantage,
enjoyed in the standard model of particle physics, of retaining the vital
features of gauge symmetries, such as Ward identities, in a future theory of
quantum gravity.

In recent work\cite{Clayton1,Clayton2}, gravitational theories based on a
bimetric structure formed from a metric and a vector field or the gradient
of a scalar field have been proposed. These theories begin with a Lorentz
and diffeomorphism invariant formulation and provide an alternative picture
to the one described here, based on a rapid change in the velocity of light
associated with a phase transition at a critical time $t=t_c$ in the early
universe when local Lorentz invariance symmetry is broken.

\vskip 0.3 true in

{\bf Acknowledgments}
\vskip 0.2 true in
This work was supported by the Natural Sciences and Engineering Research Council of Canada.
\vskip 0.5 true in

\end{document}